\documentclass[conference]{IEEEtran}
\usepackage{fancyhdr}
\usepackage{cite}
\usepackage{tabularx}
\usepackage{array}
\usepackage{makecell}
\usepackage{url}
\usepackage{amsmath,amssymb,amsfonts}
\usepackage{algorithmic}
\usepackage{graphicx}
\usepackage{textcomp}
\usepackage{booktabs}
\usepackage{xcolor}
\def\BibTeX{{\rm B\kern-.05em{\sc i\kern-.025em b}\kern-.08em
    T\kern-.1667em\lower.7ex\hbox{E}\kern-.125emX}}
\begin{document}

\title{HPL and HPL-MxP at Exascale on Aurora: Performance Engineering and Characterization}

\author{
\IEEEauthorblockN{
Kazushige Goto\IEEEauthorrefmark{1},
Huda Ibeid\IEEEauthorrefmark{1},
Kalyan Kumaran\IEEEauthorrefmark{4},
Servesh Muralidharan\IEEEauthorrefmark{4},
Anthony-Trung Nguyen\IEEEauthorrefmark{1},
Aditya Nishtala\IEEEauthorrefmark{1}
}
\IEEEauthorblockA{
\IEEEauthorrefmark{1}\textit{Intel Corporation}, Santa Clara, CA, USA\\
\IEEEauthorrefmark{4}\textit{Argonne National Laboratory}, Lemont, IL, USA
}
\IEEEauthorblockA{
\{kazushige.goto,
huda.ibeid,
anthony.d.nguyen,
aditya.nishtala\}@intel.com\\
\{kumaran,
servesh\}@anl.gov
}
}

\maketitle

\begin{abstract}
Running HPL and HPL-MxP at exascale requires more than optimized numerical kernels; it requires coordinated performance engineering across communication, resource mapping, runtime configuration, CPU--GPU interaction, and mixed-precision execution. This paper reports experience from large-scale benchmarking campaigns on Aurora, the Intel-based exascale supercomputer at the Argonne Leadership Computing Facility. Aurora progressed from 0.585~EF/s on 5,439 nodes to 1.012~EF/s on 9,234 nodes in FP64 HPL, while HPL-MxP reached 11.64~EF/s on 9,500 nodes. We describe the system-level choices behind these results and grade the evidence for each (measured, operational, or reasoned), including locality-aware rank, GPU, memory, and NIC placement; explicit CPU--GPU pipelining; host-buffer MPI and transport configuration; phase-specific NIC assignment; AMX-accelerated mixed precision; and a hybrid point-to-point/collective path introduced after synchronization stalls at scale. This experience-based study concludes with practitioner guidance separating Aurora-specific findings from broader principles for tightly coupled heterogeneous systems.

\end{abstract}


\section{Introduction}\label{Introduction}

The transition to exascale computing enables scientific simulation and data-driven workloads at unprecedented scale, but sustaining performance on heterogeneous platforms requires more than raw hardware capability. At exascale, performance emerges from coordinated performance engineering across algorithms, runtime systems, software, and hardware, while addressing challenges of scalability, resilience, and robustness at full-system scale.

Aurora~\cite{aurorapaper,auroraweb}, an exascale system deployed at the Argonne Leadership Computing Facility, exemplifies these challenges. Aurora integrates over 10,000 compute nodes, each combining Intel Xeon CPU Max Series processors with on-package high-bandwidth memory~\cite{9731107}, Intel Data Center GPU Max Series accelerators interconnected via Xe-Link~\cite{pvc22,hotchips2021blythe}, and the HPE Slingshot-11 interconnect~\cite{CUG2022_Slingshot,4556717}. This architecture represents the largest production deployment of Slingshot to date and one of the first large-scale deployments of Intel discrete GPUs, supported by a new software ecosystem based on Intel oneAPI~\cite{oneapi}. While offering substantial performance potential, this heterogeneous design introduces challenges in communication behavior, locality management, CPU--GPU interaction, and robustness at full-system scale.

In this work, we report a large-scale performance-engineering and characterization study of the High Performance Linpack (HPL) benchmark and its mixed-precision counterpart, HPL-MxP, on Aurora. We document the validated configurations and the system-level choices across communication, resource mapping, CPU--GPU overlap, and mixed-precision execution that supported exascale runs during successive large-scale benchmarking campaigns. HPL measures sustained FP64 performance and defines the TOP500 list~\cite{TOP500}, while HPL-MxP combines reduced-precision arithmetic with iterative refinement to reflect emerging scientific and AI workloads~\cite{HPLMxP}. Together, these benchmarks stress compute, memory, and interconnect subsystems at extreme scale, providing insight into both peak performance and system stability. They also provide a compact setting for studying how architectural locality, communication configuration, and runtime behavior interact at near full-system scale.

With the validated configurations, Aurora sustained 1.012~EF/s on FP64 HPL and 11.64~EF/s on mixed-precision HPL-MxP across more than 9,000 nodes, corresponding to 78.8\% parallel scaling efficiency when normalized to single-node performance. While prior work demonstrated sustained exascale performance on GPU-centric systems such as Frontier~\cite{olcf_frontier,olcf_frontier_hpl}, Aurora differs in several important architectural respects. In particular, CPUs integrate on-package high-bandwidth memory (HBM)~\cite{ibeid2025hbm} and support Advanced Matrix Extensions (AMX)~\cite{amx}; network interfaces are attached to CPUs rather than GPUs; and each node exposes more network interfaces than accelerators, enabling communication specialization. At system scale, communication is provided by the HPE Slingshot-11 interconnect organized as a high-radix dragonfly topology, which shapes collective behavior and sensitivity to transient network events.

This work does not propose new numerical algorithms for LU factorization. Instead, it reports the system-level choices used in the validated Aurora configurations and the reasoning behind them. While motivated by Aurora's design, several principles observed in this study appear relevant beyond a single system and inform performance engineering for a broad class of current and future heterogeneous HPC platforms.

This paper makes the following contributions:
\begin{itemize}
\item Reports a large-scale performance-engineering and characterization study of HPL and HPL-MxP on the Aurora supercomputer, including the progression from 0.585~EF/s on 5,439 nodes to 1.012~EF/s on 9,234 nodes for FP64 HPL, and 11.64~EF/s on 9,500 nodes for HPL-MxP.
\item Documents the validated near full-system HPL and HPL-MxP configurations on Aurora, including process grids, block sizes, rank/GPU/NIC mapping, host-buffer communication choices, transport configuration, and mixed-precision execution strategy.
\item Reports phase-level measurements from near full-system HPL runs on Aurora, quantifying how the expected shift from GEMM-dominated to communication-dominated execution manifests as the trailing matrix shrinks.
\item Describes the system-level performance-engineering choices used in the validated configurations, including locality-aware resource mapping, explicit CPU--GPU pipelining, phase-specific NIC use, AMX-accelerated low-precision CPU computation, and a hybrid P2P/collective synchronization path introduced after stalls at scale.
\item Distills practitioner guidance for heterogeneous exascale systems, distinguishing Aurora-specific observations from broader principles for locality, communication configuration, CPU--GPU overlap, collective sensitivity, and mixed-precision orchestration.
\end{itemize}

\noindent\textbf{Scope and evidence.}
This paper is an experience-based performance-engineering characterization study. The reported configurations were developed during limited large-scale allocation windows, where controlled single-variable ablations at comparable scale were generally infeasible. Accordingly, we avoid claiming isolated performance attribution for every configuration choice. Instead, we report the validated configurations, measured phase behavior from large-scale runs, operational observations from benchmarking campaigns, and system-level reasoning grounded in Aurora's architecture, runtime behavior, and prior characterization studies.

\section{Related Work}

\subsection{HPL Optimization on Exascale Systems}

As GPU-accelerated systems have become dominant at leadership scale, recent work has focused on restructuring HPL to maximize accelerator residency while mitigating latency-sensitive phases and communication bottlenecks.

Prior work describes the design and optimization of \texttt{rocHPL} for exascale accelerated nodes (e.g., Frontier), emphasizing GPU-resident trailing updates while retaining CPU panel factorization and pivoting, together with CPU time-sharing and communication-hiding strategies to improve strong scaling~\cite{olcf_frontier_hpl}. Complementary analyses study HPL performance and tuning on Frontier in depth, highlighting panel factorization and its interaction with synchronization and communication as key limiters of efficiency at extreme scale~\cite{lu2025insights}.

\subsection{Mixed-Precision Benchmarks: HPL-AI and HPL-MxP}

HPL-MxP was introduced as a mixed-precision benchmark that combines low-precision LU factorization with non-stationary iterative refinement based on GMRES, while supporting scalable input-matrix generation and addressing scaling effects on backward error~\cite{dongarra2025hplmxp}. Prior work demonstrates how FP16 tensor-core arithmetic can accelerate mixed-precision iterative refinement while preserving numerical stability, providing key motivation for mixed-precision LU and refinement–based benchmarks on GPU-accelerated systems~\cite{haidar2018tensor}.

At system scale, prior studies demonstrate that mixed-precision Linpack-style benchmarks can be scaled to systems such as Summit and Frontier, showing that communication, synchronization, and algorithmic coordination increasingly dominate as concurrency grows~\cite{lu2022climbing}. Beyond GPU-centric platforms, large-scale mixed-precision Linpack performance has also been demonstrated on CPU-based systems such as Fugaku, emphasizing architecture-aware optimizations and performance considerations distinct from accelerator-focused approaches~\cite{kudo2020fugakuHPLAI}.

Several capabilities used in this work, including Aurora-tuned MPICH, libfabric/CXI, high-radix collectives, PMIx/PALS startup, and oneMKL kernels, are inherited from the Aurora software environment. The contribution of this paper lies in the configuration and characterization of how these components were combined with application-specific choices for HPL and HPL-MxP at scale, including host-buffer MPI, deterministic resource mapping, phase-specific NIC use, hybrid P2P/collective panel communication, and mixed-precision CPU--GPU orchestration.
\section{Background}\label{Background}

\subsection{Benchmarks}

\subsubsection{HPL}

High Performance Linpack (HPL) solves a dense linear system using LU factorization with partial pivoting and reports sustained FP64 performance~\cite{hplpaper,petitet2001hpl}. Following the standard HPL definition~\cite{hplpaper,petitet2001hpl}, the operation count is
\begin{equation*}
    \text{FLOP}_\text{total} = \tfrac{2}{3}N^{3} + \mathcal{O}(N^{2}),
\end{equation*}
and the reported performance is
\begin{equation*}
    R_\text{max} = \frac{\tfrac{2}{3}N^{3} + \mathcal{O}(N^{2})}{\text{Runtime}}.
\end{equation*}

Here, \(N\) is the order of the linear system and \(R_\text{max}\) is the maximal sustained performance achieved, in floating-point operations per second (FLOP/s), which is the value reported for the TOP500 list.

Parallel HPL uses a two-dimensional block-cyclic distribution over a \(P \times Q\) process grid with block size \(N_B\). Each panel step consists of panel factorization (PFACT), panel broadcast (BCAST), row swapping (SWAP), triangular solve (DTRSM), and trailing matrix update (DGEMM). DGEMM dominates the floating-point work for large problems, while PFACT, BCAST, and SWAP expose latency and communication costs at scale.

At panel step \(k\) with remaining trailing dimension \(n_k\), the per-step phase times scale as
\begin{align*}
T_{\text{DGEMM}} &\sim \frac{n_k^{2}\, N_B}{R_{\text{DGEMM}}}, \\
T_{\text{PFACT}} &\sim \frac{n_k\, N_B^{2}}{R_{\text{PFACT}}}, \\
T_{\text{BCAST}},\, T_{\text{SWAP}} &\sim \frac{n_k\, N_B}{BW_{\text{net}}}.
\end{align*}

Thus, in a simplified non-overlapped view,
\begin{equation*}
T_{\text{total}} \approx
   \sum_k
   \max\!\bigl(
       T_{\text{DGEMM}},\;
       T_{\text{PFACT}} + T_{\text{BCAST}},\;
       T_{\text{SWAP}}
   \bigr).
\end{equation*}

For sufficiently large problem sizes, early iterations are dominated by DGEMM, while later iterations increasingly expose communication- and latency-sensitive phases as the trailing matrix shrinks. We use this model qualitatively, not as a calibrated runtime predictor, to interpret the measured phase behavior in Section~\ref{Results}.

On Aurora, DGEMM executes on GPUs, while panel factorization, row swaps, triangular solves, MPI coordination, and GPU orchestration are managed by CPUs. Sustained performance therefore depends on CPU--GPU pipelining, communication placement, and topology-aware process mapping.

\subsubsection{HPL-MxP}

HPL-MxP is a mixed-precision counterpart to HPL designed to reflect emerging AI and scientific applications~\cite{HPLMxP}. It accelerates the solution of dense linear systems by executing most arithmetic operations in reduced precision, with iterative refinement used to recover double-precision accuracy. While the benchmark retains the broad structure of HPL, factorizations and trailing matrix updates are performed in low precision (FP32, FP16, or BF16), whereas residual evaluation and corrections are computed in double precision (FP64).

In the Aurora implementation studied here, matrices are stored in BF16 format, substantially reducing memory requirements and enabling larger problems to fit in GPU memory. Although reduced precision may increase the number of iterative refinement steps, the resulting gains in GPU throughput and utilization outweigh this additional cost. On Aurora, GPUs perform the reduced-precision factorization and trailing matrix updates using BF16-optimized kernels, while CPUs compute residuals and apply refinement in double precision, enabling efficient utilization of both CPUs and GPUs.

\subsection{Aurora Architecture}

\begin{figure}[t]
    \centering
    \includegraphics[width=0.95\linewidth]{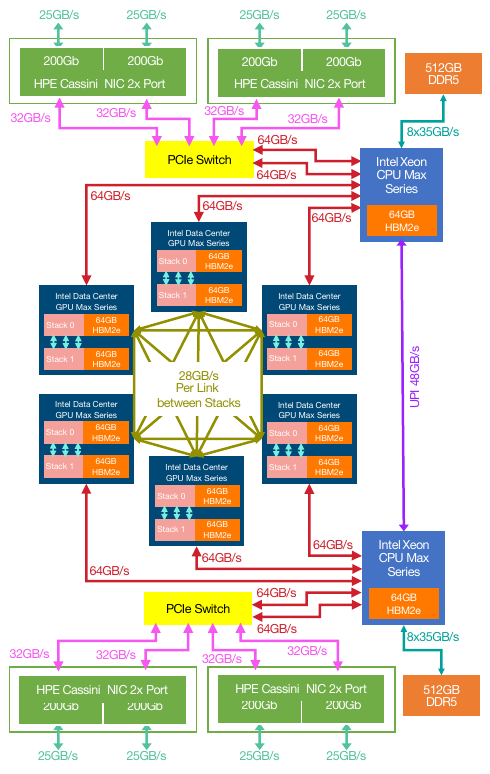}
    \caption{Aurora Exascale Compute Blade (ECB).}
    \label{fig:aurora-node}
\end{figure}

Aurora consists of 10,624 compute nodes connected by the HPE Slingshot-11 interconnect in a dragonfly topology~\cite{CUG2022_Slingshot,4556717}. Each Exascale Compute Blade (ECB), shown in Figure~\ref{fig:aurora-node}, contains two Intel Xeon CPU Max Series processors and six Intel Data Center GPU Max Series accelerators~\cite{9731107,pvc22,hotchips2021blythe}. Each CPU integrates 64~GB of HBM2e and 512~GB of DDR5 memory, and supports Intel AMX for low-precision matrix operations~\cite{ibeid2025hbm,amx}. Each CPU connects to three GPUs over PCIe Gen5 x16 links, while Xe-Link connects the six GPUs.

Network connectivity is provided by eight HPE Cassini NICs per node, attached through PCIe switches. This CPU-mediated, multi-NIC node topology is central to the resource-mapping and communication choices described in Section~\ref{sec:Optimizations}: rank, memory, GPU, and NIC placement must preserve locality, and different communication phases can be mapped to different NICs.

\section{Experimental Setup}\label{Setup}

\subsection{Benchmark Parameters}

Table~\ref{tab:benchmark-config} summarizes the configurations used for the largest-scale runs. Parameters such as block size (\(N_B\)) and process grid dimensions (\(P \times Q\)) are tuned to balance GPU arithmetic intensity, communication overhead, and overlap between CPU and GPU tasks. For both HPL and HPL-MxP, we select the largest \(N_B\) values that fit in GPU memory. Because HPL-MxP employs BF16 matrix storage rather than FP64, it enables block sizes up to four times larger than HPL. Problem sizes (\(N\)) are chosen to saturate available memory, while process grid dimensions are selected to maintain balanced communication patterns at scale.

HPL employs 6 MPI ranks per node (PPN=6), with each rank bound to a GPU/NIC pair, while HPL-MxP employs 2 MPI ranks per node (PPN=2), with each rank managing 3 GPUs through a thread-based runtime. This mapping reflects the distinct communication and computation characteristics of the two benchmarks. Further discussion of tuning trade-offs is provided in Section~\ref{sec:Optimizations}.

\begin{table}[ht]
\caption{Benchmark configurations for large-scale runs on Aurora.}
\label{tab:benchmark-config}
\centering
\begin{tabular}{|l|l|l|}
\hline
\textbf{Parameter} & \textbf{HPL} & \textbf{HPL-MxP} \\
\hline
Nodes                       & 9,234              & 9,500               \\
Processes per node (PPN)    & 6                  & 2                   \\
Process grid \(P \times Q\) & \(162 \times 342\) & \(152\times 125\)   \\
Block size \(N_B\)           & 384                & 1,536               \\
Problem size \(N\)          & 28,773,888         & 57,693,696          \\
Precision                   & FP64               & BF16 + FP32 + FP64  \\
Look-ahead depth            & 1                  & 1                   \\
\hline
\end{tabular}
\end{table}

\subsection{Performance Metrics}\label{sec:metrics}
We report performance using parallel scaling efficiency, which captures how effectively the system scales with increasing node count. Parallel scaling efficiency is defined as
\[
\eta_{\mathrm{scale}} = \frac{R_{\text{max}}(N)}{N \cdot R_{\text{max}}(1)},
\]
where $R_{\text{max}}(N)$ is the achieved performance on $N$ nodes and $R_{\text{max}}(1)$ is the achieved performance on a single node. This metric reflects the efficiency with which performance scales as the system size increases, independent of theoretical peak hardware capability. For the runs in this paper, $ R_{max}(1) = 139 $ TF/s, measured on a single node (PPN = 6, NB = 384, N = 307968). Because a single-node run has no inter-node BCAST or SWAP traffic and a smaller N per node, $ \eta_{scale} $ folds together network cost, load imbalance, and problem-size effects; we use it as a system-level indicator, not as a communication-only metric.

\section{System-Level Performance Engineering}\label{sec:Optimizations}

Sustaining exascale HPL and HPL-MxP runs on Aurora requires coordinated choices spanning algorithmic parameters, software and runtime configuration, and hardware-aware execution. These choices were tuned jointly during large-scale benchmarking campaigns. Because isolated ablations at comparable scale were infeasible, this section focuses on the validated configuration, the rationale for each choice, and the supporting evidence from phase measurements, operational observations, and architectural constraints. Table~\ref{tab:opt-impact} summarizes these choices alongside the problem each addresses, its intended effect, and a graded indication of the supporting evidence, ranging from measurements reported in this paper, to operational observations from large-scale runs, to architecture- and mechanism-based reasoning.

\subsection{Algorithmic-Level Choices}\label{sec:AlgorithmicLevelChoices}

Algorithmic decisions determine arithmetic intensity, memory footprint, and the balance between CPU and GPU workloads. In Aurora’s HPL and HPL-MxP implementations, these choices center on storage format, block size, and process mapping. HPL is restricted to FP64 storage, which constrains memory capacity, while HPL-MxP permits reduced-precision storage. In both cases, the problem size is chosen to saturate memory, and the MPI process grid is tuned to balance communication across Aurora’s Slingshot-11 topology (Table~\ref{tab:benchmark-config}).

In HPL, CPUs perform panel factorization (PFACT), row swaps (SWAP), MPI communication, and control logic, while GPUs accelerate DGEMM computations. Intel’s OpenCL runtime manages GPU kernel launches and CPU--GPU data transfers, enabling pipelined overlap of CPU-managed tasks with GPU kernels. CPU operations rely on single-threaded Intel oneMKL BLAS, while GPU computations use a modified oneMKL DGEMM kernel tuned for Aurora. HPL-MxP retains this overall structure but replaces FP64 computation with a mixed-precision approach in which CPU-based iterative refinement restores FP64 accuracy.

The storage format directly shapes feasible block sizes (\(N_B\)) and overall problem size (\(N\)) because both are bounded by GPU memory capacity. With FP64 storage, HPL quickly exhausts memory, forcing smaller \(N_B\) values and limiting \(N\). HPL-MxP instead stores the matrix in BF16, reducing the footprint by up to \(4\times\). This relaxation enables substantially larger block sizes and nearly \(2\times\) larger problems. Larger blocks increase arithmetic intensity, while reduced-precision computation delivers higher throughput on both GPUs and CPUs, sustaining performance at scale.

The two benchmarks adopt different process layouts to balance computation and communication. HPL uses 6 MPI ranks per node (PPN=6), each bound to a dedicated GPU/NIC pair, maximizing per-rank bandwidth and reducing contention. This configuration supports communication-intensive panel factorizations alongside GPU updates and provides high aggregate bandwidth. In contrast, HPL-MxP uses 2 MPI ranks per node (PPN=2), with each rank managing 3 GPUs through a thread-based runtime. With larger block sizes and no pivoting, HPL-MxP incurs lower communication overhead but increases the cost of CPU-based TRSM, as larger blocks make TRSM solves more expensive. A coarser process mapping alleviates this bottleneck by reducing CPU contention and improving TRSM throughput, while the reduced communication requirements of HPL-MxP make the bandwidth trade-off acceptable.

\begin{table*}[t]
\centering
\caption{System-level choices in the validated Aurora HPL/HPL-MxP configuration. Evidence strength is graded as \emph{Measured} (phase or scaling measurements in this paper), \emph{Operational} (observed during large-scale runs, not a controlled test), or \emph{Reasoned} (architecture- or mechanism-motivated). Applicability reflects the mechanism's dependence on architecture, not the strength of evidence; the two broad entries are reasoned, not measured (see Sec. \ref{sec:practitionerguidance}).}
\label{tab:opt-impact}
\footnotesize
\setlength{\tabcolsep}{3pt}
\renewcommand{\arraystretch}{1.25}
\begin{tabular}{|
>{\raggedright\arraybackslash}p{0.05\textwidth}|
>{\raggedright\arraybackslash}p{0.15\textwidth}|
>{\raggedright\arraybackslash}p{0.23\textwidth}|
>{\raggedright\arraybackslash}p{0.23\textwidth}|
>{\raggedright\arraybackslash}p{0.10\textwidth}|
>{\raggedright\arraybackslash}p{0.15\textwidth}|}
\hline
\textbf{Sec.} & \textbf{Choice} & \textbf{Problem addressed} & \textbf{Intended effect} & \textbf{Evidence} & \textbf{Expected applicability} \\ \hline

V-A &
Process grid (P$\times$Q) tuning &
Communication overtakes compute as the trailing matrix shrinks &
Extend the GEMM-dominated region; keep execution compute-bound longer &
Measured &
Benchmark-specific \\ \hline

III-A/V-A &
BF16 storage + iterative refinement &
FP64 exhausts GPU memory, limiting $N$ and $N_B$ &
Cut footprint up to $4\times$; larger blocks and problems &
Reasoned &
HPL-MxP / general \\ \hline

V-A &
Ranks per node (HPL PPN=6, HPL-MxP PPN=2) &
Balance communication vs.\ CPU-side TRSM and refinement cost &
Match per-node parallelism to each benchmark's bottleneck &
Reasoned &
Benchmark-specific \\ \hline

V-B-1 &
Host-buffer MPI, GPU-aware paths disabled &
Both codes communicate from host memory &
Remove unused GPU-buffer handling and monitoring &
Reasoned &
Host-buffer codes \\ \hline

V-B-2 &
Enlarged CQs + deterministic NIC mapping &
CQ overflow and run-to-run jitter at scale &
Provision transport capacity; stabilize NIC selection &
Operational &
Aurora / Slingshot \\ \hline

V-B-3 &
Hybrid P2P/collective panel path &
Collective stalls amplify localized network delays &
Limit propagation of synchronization stalls &
Operational &
HPL-like synchronization \\ \hline

V-C-1 &
Rank/GPU/memory/NIC co-location &
Cross-socket and remote-NUMA traffic &
Preserve locality across compute, memory, and NICs &
Reasoned &
Broad \\ \hline

V-C-2 &
Multiple OpenCL queues and copy engines &
PCIe latency stalls the CPU--GPU pipeline &
Overlap transfers with computation &
Reasoned &
Broad \\ \hline

V-C-3 &
Phase-specific SWAP/BCAST NIC assignment &
SWAP and BCAST contend on shared NICs &
Separate latency- and bandwidth-sensitive traffic &
Reasoned &
Multi-NIC nodes \\ \hline

V-C-4 &
AMX-accelerated BF16 GEMM (HPL-MxP) &
CPU-side refinement and low-precision work on critical path &
Raise CPU-side GEMM throughput &
Operational &
AMX-capable CPUs \\ \hline

\end{tabular}
\end{table*}

\subsection{Software and Runtime Configuration}\label{sec:SoftwareOptimizations}

Software and runtime choices span both system-level configuration and application-level communication strategies. At the system level, MPI configuration and transport-layer parameters determine the efficiency, robustness, and stability of collective communication at extreme node counts. At the application level, resilience mechanisms mitigate transient network instabilities that can disrupt multi-hour runs at scale.

\subsubsection{MPI Configuration and Capabilities}

MPI runtime configuration is a key factor in scalability. Aurora's MPI environment is based on MPICH~\cite{doi:10.1177/10943420241311608} and is configured for the HPE Slingshot/Cassini interconnect~\cite{mpich-sc17}. The Aurora MPICH stack provides high-radix collectives, scalable startup through PMIx/PALS, Yaksa datatype support, and GPU-aware communication mechanisms~\cite{doi:10.1177/10943420241311608,mpich-sc17,mpich-high-radix21,yaksa,pmix,pals}. Because HPL and HPL-MxP communicate through host-resident buffers managed by the CPUs, GPU-aware MPI support was disabled in the validated configuration.

\subsubsection{Transport Layer Configuration -- libfabric/cxi}

Beyond MPI, transport-layer configuration strongly influences scalability. Aurora employs the \texttt{libfabric} communication stack with the \texttt{cxi} provider for its Slingshot-11 interconnect. At scale, performance depends on how the transport layer is tuned to handle high message rates and overall communication volume. The runtime is configured following three guiding principles: provision sufficient queue resources, enforce deterministic mapping, and minimize monitoring overhead.

Following these principles, the runtime configuration is adjusted as follows.
\begin{itemize}
    \item \textbf{Provisioning resources:} Ensures sufficient transport capacity for collective-heavy workloads.  
    \begin{itemize}
        \item \verb|FI_PROVIDER=cxi| selects the CXI provider. 
        \item \verb|FI_CXI_DEFAULT_CQ_SIZE=131072| enlarges completion queues to prevent CQ overflows during high-volume collective operations.
    \end{itemize}
    
    \item \textbf{Enforcing deterministic mapping:} Provides stable NIC assignment and minimizes run-to-run jitter.  
    \begin{itemize}    
        \item \verb|MPIR_CVAR_ODD_EVEN_CLIQUES=1| improves logical rank ordering for balanced utilization. 
        \item \verb|MPIR_CVAR_CH4_OFI_ENABLE_STRIPING=0| disables endpoint striping. 
        \item \verb|MPIR_CVAR_CH4_OFI_ENABLE_HASHING=0| disables hashing-based endpoint selection.
    \end{itemize}
    
    \item \textbf{Reducing monitoring overhead:} Removes unnecessary polling and registration costs when GPU-aware MPI is not enabled.  
    \begin{itemize}        
        \item \verb|FI_MR_CACHE_MONITOR=memhooks| leverages lightweight memory hooks for MR cache coherence.
        \item \verb|FI_MR_ZE_CACHE_MONITOR_ENABLED=0| disables Level Zero MR monitoring. 
        \item \verb|MPIR_CVAR_ENABLE_GPU=0| explicitly confirms host-only communication.
    \end{itemize}
\end{itemize}

\subsubsection{Communication Resilience -- Handling Flap Events}

Large-scale validation experience on Aurora indicates that transient network events and associated MPI-level retries can occur at near full-system scale, particularly for collective-heavy workloads~\cite{scalingmpiaurora}. Because collectives such as \texttt{MPI\_Allreduce} impose global synchronization, a delay affecting a single rank can stall an entire process column. Such stalls are particularly consequential during communication-sensitive panel operations, where a localized delay can propagate through subsequent computation.

To reduce sensitivity to these stalls, the implementation includes a hybrid reduction path that uses point-to-point (P2P) communication for the first panel-column reduction and optimized collectives for subsequent columns. The first panel-column reduction uses direct P2P transfers instead of \texttt{MPI\_Allreduce} because it is particularly sensitive to synchronization delays. Subsequent reductions use optimized collective operations to retain their performance benefits. Although the P2P path introduces additional communication rounds, the hybrid path was used in the validated configuration after large-scale operational observations indicated reduced sensitivity to synchronization stalls. This design limits the extent to which a localized delay in a synchronization-sensitive panel operation propagates through the computation. The added cost is bounded: the P2P path applies only to the first panel-column reduction, which is $O(NB \cdot \log P)$ messages per panel step versus one collective.

\subsection{Hardware-Aware Configuration}\label{sec:HardwareOptimizations}

Hardware-aware optimizations exploit Aurora’s architectural features, targeting NUMA locality, PCIe transfer concurrency between CPUs and GPUs, NIC assignment, and Intel AMX instructions for accelerating low-precision compute.

\subsubsection{NUMA--Aware Process Mapping}

Each MPI process is mapped such that its CPU cores, GPUs, NIC, and memory allocations are co-located on the same CPU, preserving locality across computation, communication, and memory accesses. Each Aurora CPU exposes two NUMA nodes, one corresponding to HBM2e and the other to DDR5. For large-scale HPL runs, the global matrix resides in DDR due to capacity constraints, as problem sizes exceed the available HBM capacity. DDR is therefore used as the primary memory for CPU-managed data structures and bulk storage.

HBM2e is used during panel factorization, where tiles required for TRSM and row-exchange operations are staged from DDR into HBM to benefit from higher bandwidth. While this supports efficient execution of CPU-side panel operations, overall runtime is dominated by GPU-accelerated GEMM.

CPU affinity is enforced with \texttt{taskset} or \texttt{numactl}, memory placement with \texttt{numactl} (e.g., \texttt{--membind}), and GPU binding with device-affinity masks (e.g., \texttt{ZE\_AFFINITY\_MASK}). This mapping ensures that CPU execution, GPU operations, memory accesses, and NIC traffic remain within the same CPU-local memory domain, which reduces UPI traffic between CPUs. By minimizing remote memory access and intra-node contention, the configuration is intended to lower latency and provide more predictable performance for CPU-managed operations.

\subsubsection{Overlapping CPU--GPU PCIe Data Transfers with Computation}

On Aurora, HPL and HPL-MxP are executed as a pipelined sequence of GPU kernels, device-to-host (D2H) transfers, CPU computation with MPI communication, host-to-device (H2D) transfers, and subsequent GPU kernels. Without careful scheduling, PCIe latency would stall this pipeline and reduce GPU utilization. To mitigate this, the benchmarks explicitly overlap CPU--GPU data transfers with computation. This strategy exploits the availability of multiple DMA copy engines on Aurora GPUs, allowing communication and computation to proceed in parallel.

Concurrency is achieved through explicit queue management. Within each process, the benchmarks manage OpenCL command queues directly. Each queue issues a single operation at a time, and concurrency is enabled by issuing independent transfers and kernels across multiple compute and copy queues. Panel data is transferred from GPU to CPU (D2H) in parallel across multiple copy engines, while CPU-to-GPU (H2D) updates are concurrently scheduled on available copy engines to sustain bidirectional throughput. OpenCL events synchronize these operations so that kernels execute on already-resident data while upcoming transfers progress in the background.

This design supports a continuous execution pipeline by allowing transfers and kernels to proceed concurrently. In the validated configuration, this overlap was used to reduce exposure to PCIe latency and maintain GPU progress during CPU-managed phases.

\subsubsection{Phase--Specific NIC Assignment}

Each Aurora node includes eight NICs (four per CPU). To reduce contention and preserve locality, HPL and HPL-MxP assign NICs to communication phases in a phase-specific manner. Row swapping operations (SWAP), which are latency-sensitive and occur on every panel factorization, are bound to the first NIC (NIC 0) on each CPU, while broadcasts (BCAST), which are network bandwidth-sensitive, are distributed across the remaining three NICs (NICs 1--3). This separation isolates SWAP traffic from BCAST traffic, reducing contention within the same CPU.

Binding is enforced at communicator creation through MPICH \texttt{multi\_nic\_pref\_nic} hints, which provide explicit control over NIC selection. This mapping provides consistent NIC usage across runs and separates latency-sensitive SWAP traffic from bandwidth-sensitive BCAST traffic, which is intended to reduce contention during the most communication-intensive phases of HPL and HPL-MxP.

\subsubsection{Exploiting Intel AMX for Low-Precision Compute}\label{sec:ExploitingIntelAMX}

Aurora’s Xeon Max CPUs support Advanced Matrix Extensions (AMX), which provide hardware acceleration for low-precision matrix operations such as \texttt{BF16} GEMMs~\cite{amx}. In HPL-MxP, AMX is leveraged in two contexts. It accelerates CPU-based iterative refinement and low-precision panel factorization. Both stages can benefit from higher throughput on GEMM operations, reducing CPU-side work in the mixed-precision execution path.

Intel oneMKL dispatches AMX-accelerated kernels where supported, allowing the benchmark to use these hardware features without additional application-level code changes. AMX enablement was the most clearly identifiable configuration change associated with the HPL-MxP improvement between the two campaigns reported in Table~\ref{tab:campaign-progression}.

\subsection{Practitioner Guidance}\label{sec:practitionerguidance}

From these large-scale benchmarking campaigns, we identify several practices that may transfer beyond HPL and HPL-MxP. Items 1, 4, and 6 are reasoned from mechanism and were not isolated experimentally; items 2, 3, and 5 rest on operational observations on Aurora (Table \ref{tab:opt-impact}). We present them as working hypotheses for other systems, not as validated rules.

\begin{enumerate}
\item \textbf{Co-locate communicating resources.} On multi-socket heterogeneous nodes, rank, CPU-core, memory, GPU, and NIC placement should be treated as a single locality-aware mapping problem rather than tuned independently.

\item \textbf{Match MPI features to the actual buffer path.}
HPL and HPL-MxP use host-resident communication buffers, so GPU-aware MPI paths that are not exercised can be disabled. Applications communicating directly from GPU memory may instead benefit from enabling GPU-aware paths.

\item \textbf{Make NIC selection explicit on multi-NIC nodes.}
Applications with distinct latency-sensitive and bandwidth-sensitive phases may benefit from deterministic NIC assignment or communicator-level NIC hints.

\item \textbf{Pipeline CPU work, GPU kernels, transfers, and communication.}
When host-mediated communication lies on the critical path, relying on implicit runtime overlap may be insufficient, and explicit pipelining of transfers, kernels, and communication may be required.

\item \textbf{Treat large collectives as operational risk points.}
At extreme scale, tightly synchronized collectives can amplify localized delays. Applications with known synchronization-critical phases may benefit from algorithm-aware fallback paths.

\item \textbf{Tune mixed precision end to end.}
Reduced precision changes memory footprint, block size, process layout, CPU--GPU balance, and refinement cost; it is not only a kernel-level change.
\end{enumerate}

The specific environment variables, AMX behavior, and NIC topology described here are Aurora-specific, whereas the broader principles, locality-aware mapping, communication-path matching, pipeline overlap, collective sensitivity, and end-to-end mixed-precision orchestration, are likely to apply to other tightly coupled heterogeneous systems.
\section{Performance Results and Analysis}\label{Results}

Building on the system-level choices described in Section~\ref{sec:Optimizations}, we now present performance results obtained on Aurora at near full-system scale. Table~\ref{tab:campaign-progression} summarizes the progression of HPL and HPL-MxP results across three benchmarking campaigns: SC23 corresponds to the initial large-scale HPL run, ISC24 to near full-system HPL and HPL-MxP runs, and SC24 to the HPL-MxP result associated with AMX enablement. We present results for FP64 HPL and mixed-precision HPL-MxP achieved by the validated combination of algorithmic, software, runtime, and hardware-aware choices, and conclude with a discussion of challenges encountered at this scale.

\begin{table}[h]
\caption{HPL and HPL-MxP progression across Aurora campaigns.}
\label{tab:campaign-progression}
\centering
\footnotesize
\setlength{\tabcolsep}{3pt}
\begin{tabular}{|l|cc|cc|}
\hline
\textbf{Campaign} &
\multicolumn{2}{c|}{\textbf{HPL}} &
\multicolumn{2}{c|}{\textbf{HPL-MxP}} \\
\cline{2-5}
 & \textbf{Nodes} & \textbf{EF/s} & \textbf{Nodes} & \textbf{EF/s} \\
\hline
Nov 2023 (SC23)  & 5,439 & 0.585 & --    & --    \\
May 2024 (ISC24) & 9,234 & 1.012  & 9,234 & 10.6  \\
Nov 2024 (SC24)  & 9,234 & 1.012  & 9,500 & 11.64 \\
\hline
\end{tabular}
    \par
    \medskip
    \begin{minipage}{0.95\columnwidth}
        \footnotesize
        No further full-system HPL/HPL-MxP submissions were made after November 2024, as Aurora entered production operation and full-system allocations were redirected to science campaigns; the configuration described here remains the reference configuration on the system (as of Sep 2026).
    \end{minipage}
\end{table}

\subsection{HPL: Sustained FP64 Exascale Performance}

\begin{figure}[t]
\centering
\includegraphics[width=0.98\linewidth]{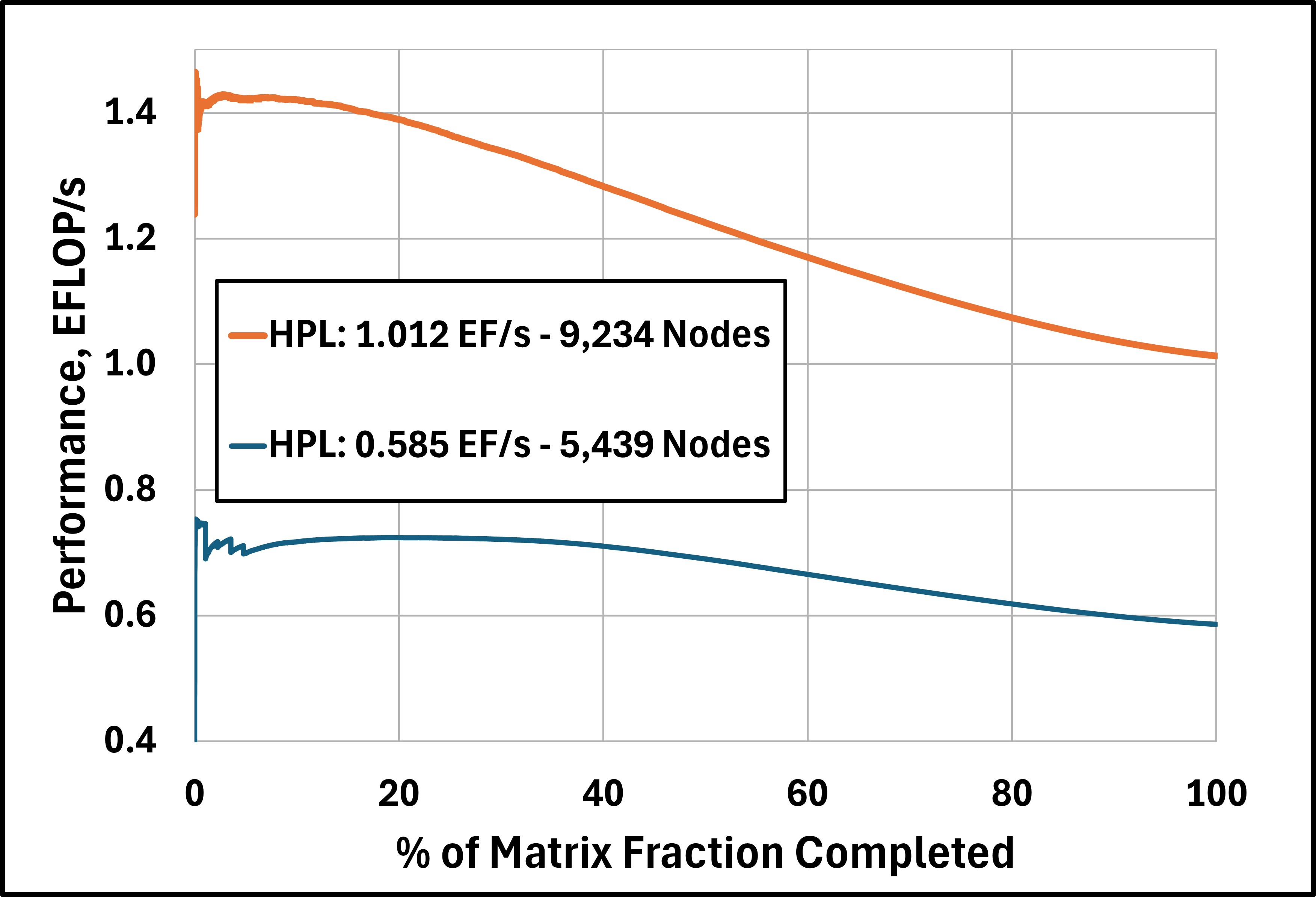}
\caption{HPL performance on Aurora, scaling from 5,439 to 9,234 nodes with sustained exascale performance.}
\label{fig:HPL_Performance}
\end{figure}

Aurora achieved 1.012~EF/s on 9,234 nodes, corresponding to 78.8\% parallel scaling efficiency relative to single-node performance. Compared to earlier large-scale experiments, efficiency was 77.3\% at 5,439 nodes and 78.8\% at 9,234 nodes. These results show that Aurora maintained high per-node throughput even at near full-system size, as shown in Figure~\ref{fig:HPL_Performance}.

\begin{figure}[ht]
\centering
\includegraphics[width=0.98\linewidth]{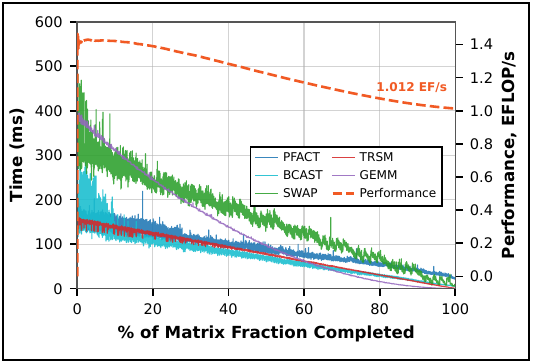}
\caption{Phase breakdown of the exascale HPL run on 9,234 nodes. GEMM dominates early, while SWAP becomes dominant as the trailing submatrix shrinks; the crossover aligns with the decline in the performance curve.}
\label{fig:hpl_breakdown}
\end{figure}

The phase breakdown of the 9,234-node exascale run is shown in Figure~\ref{fig:hpl_breakdown}. The x-axis shows progress through the LU factorization, while the left y-axis reports per-phase time in milliseconds and the right y-axis reports performance in GF/s. GEMM dominates the early portion of the run, keeping the benchmark compute-bound during the initial stages. As factorization progresses and the trailing submatrix shrinks, per-step GEMM time decreases. SWAP begins to dominate runtime once GEMM drops, marking the transition from compute-bound to communication-bound execution.

The system transitions from compute-bound to communication-bound execution when the GEMM curve falls below the SWAP curve. This crossover aligns with the drop in sustained performance indicated by the dashed orange line. Consistent with Section~\ref{sec:Optimizations}, the process grid \((P, Q)\) was tuned to extend the GEMM-dominated region and keep the run compute-bound as long as possible, while the communication and runtime choices reduce exposure to communication bottlenecks once the trailing matrix shrinks.

This crossover is consistent with the model in Section~\ref{Background}, where the DGEMM term falls faster than the panel and swap terms as the trailing matrix shrinks.

\subsubsection*{Comparison with Frontier}
Frontier's HPL campaign~\cite{olcf_frontier_hpl,lu2025insights} and Aurora's differ in where the latency-sensitive work lands. On Frontier, NICs attach to GPUs and rocHPL keeps the trailing update GPU-resident while time-sharing the CPU for panel factorization; the reported limiter at scale is panel factorization and its synchronization~\cite{lu2025insights}. On Aurora, NICs attach to CPUs, each rank owns a dedicated GPU/NIC pair, and the measured limiter as the trailing matrix shrinks is SWAP (Fig.~\ref{fig:hpl_breakdown}) rather than PFACT, consistent with host-mediated row exchange over more NICs than GPUs. On the November 2024 TOP500 list~\cite{TOP500}, Frontier reached 1.353 EF/s (65.8\% of its 2.056 EF/s Rpeak) and Aurora 1.012 EF/s (51.1\% of its 1.980 EF/s Rpeak); part of Aurora's gap reflects that CPU FP64 peak, included in Rpeak, is intentionally left to orchestration and panel work (Sec.~\ref{sec:AlgorithmicLevelChoices}). In mixed precision, Aurora's 11.64 EF/s exceeded Frontier's 11.4 EF/s on the same list~\cite{HPLMxP}, which we attribute in part to BF16 storage and CPU-side AMX refinement (Sec.~\ref{sec:ExploitingIntelAMX}).

\subsection{HPL-MxP: Mixed-Precision Acceleration}

Aurora sustained 11.64~EF/s on 9,500 nodes with HPL-MxP, representing an 11.5$\times$ speedup over FP64 HPL. At ISC24, Aurora delivered 10.6~EF/s, and by SC24, the sustained performance improved to 11.64~EF/s. The most clearly identifiable configuration change between ISC24 and SC24 was the use of AMX units on the CPUs for low-precision CPU-side computation.

Normalizing by node count, per-node HPL-MxP throughput rose from 1.148 PF/s (ISC24, 9,234 nodes) to 1.225 PF/s (SC24, 9,500 nodes), i.e., +6.7\% per node, with the remaining +2.9\% of the 9.8\% aggregate gain due to node count. Because the two campaigns also differed in allocation and system state, this per-node delta is an upper bound on the AMX contribution rather than an isolated attribution.

The LU factorization employs mixed-precision arithmetic (BF16/FP32) on GPUs to increase arithmetic intensity, followed by FP64 iterative refinement on CPUs to restore accuracy. AMX acceleration improves the expected throughput of CPU-side GEMMs in both refinement and low-precision factorization, reducing the time spent in CPU-managed work.

\begin{figure}[t]
\centering
\includegraphics[width=0.98\linewidth]{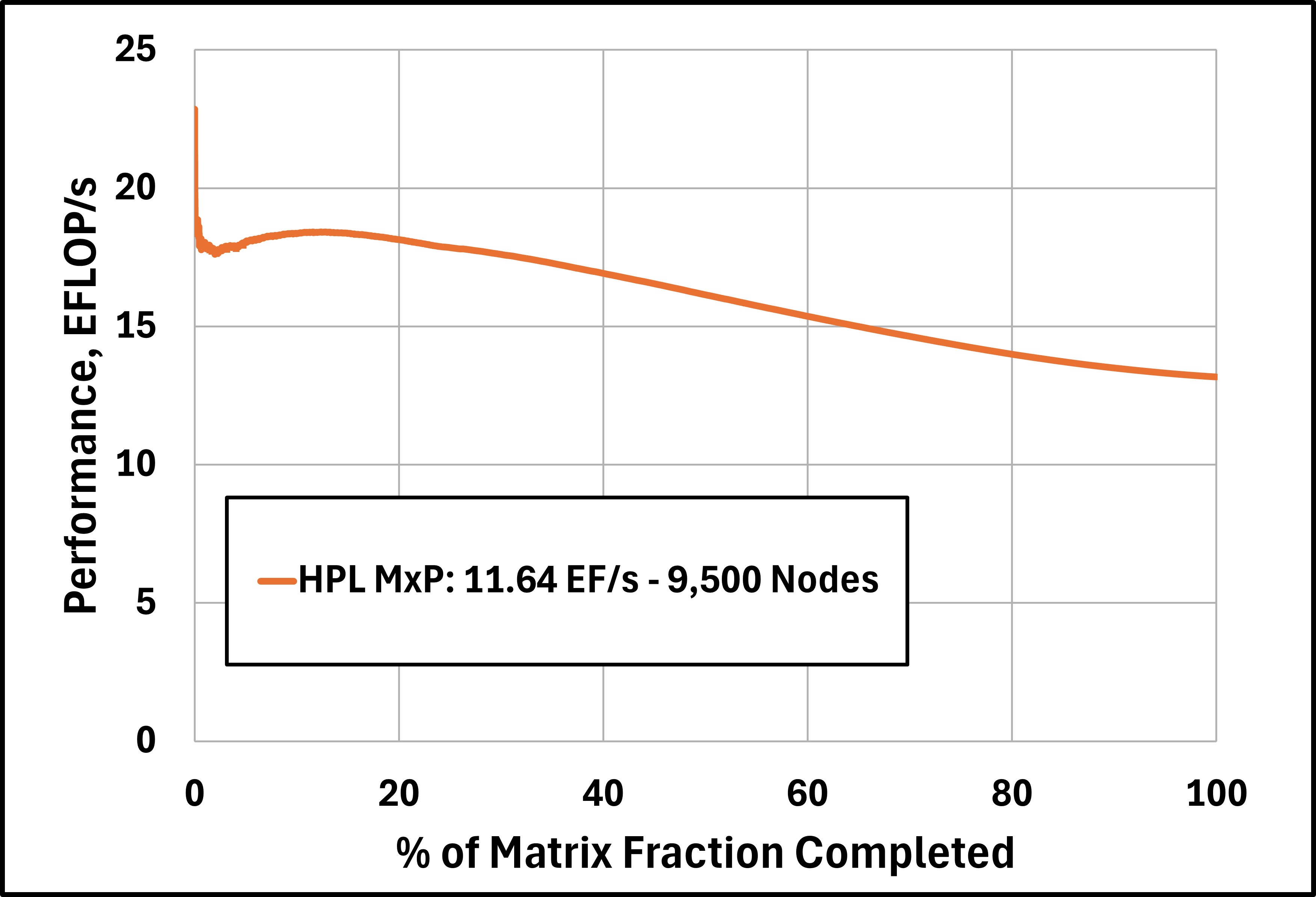}
\caption{HPL-MxP performance on Aurora at 9,500 nodes. A short warm-up is followed by a long high-performance plateau and a gradual decline as the factorization progresses, reaching a final performance of 11.64~EF/s.}
\label{fig:HPL_MXP_Performance}
\end{figure}

As shown in Figure~\ref{fig:HPL_MXP_Performance}, the per-step performance profile exhibits a brief warm-up transient at the start, sustains high performance across the early stages, and then declines gradually as the trailing submatrix shrinks and communication becomes a larger fraction of the work.

Together, these results show that Aurora sustained exascale throughput in FP64 HPL and more than an order-of-magnitude higher reported throughput in the distinct mixed-precision HPL-MxP benchmark mode. The validated combination of algorithmic, software, runtime, and hardware-aware choices supported stable performance across hours-long runs at near full-system scale. Although system-wide reliability ultimately depends on the underlying hardware, these runs show that, when stable resources were available, the configuration delivered exascale HPL and HPL-MxP performance.

\subsection{Challenges}

Despite sustained exascale performance on Aurora, large-scale runs face challenges beyond application-level choices. These challenges stem from the extreme scale of the system, where millions of components must run in synchrony for hours. Two factors are especially critical: \textit{system reliability}, which determines whether jobs can complete without interruption, and \textit{network variability}, which influences how predictably performance is delivered across thousands of nodes.

\subsubsection{System Reliability at Scale}

At exascale, reliability is strained by the sheer number of components operating concurrently. Even with robust hardware, faults such as node failures, memory errors, or storage issues become more probable during hours-long large-scale benchmarking runs. For tightly coupled applications like HPL and HPL-MxP, a single fault can cascade to job termination.

Aurora addresses these risks with hardware-level RAS features, continuous monitoring, and automated in-field repair (IFR). The job scheduler and system management software further isolate failing components and recycle healthy resources, improving overall availability.

Nevertheless, when millions of devices interact at scale, occasional failures are unavoidable. While system- and application-level mechanisms can reduce their impact, the risk of node loss or job termination remains an inherent challenge for sustained exascale jobs.

\subsubsection{Network Variability}

At near full-system scale, HPL and HPL-MxP runs are sensitive to transient network events. Aurora's Slingshot-11 interconnect provides adaptive routing and congestion management, which generally sustain high throughput across thousands of nodes. However, brief instabilities, such as flap events or localized congestion, can manifest as timeouts and retries at the MPI layer. When amplified by collective operations, these disruptions appear as short stalls or increased variability during communication-bound phases.

To mitigate such effects, large-scale jobs are preceded by validation tests~\cite{scalingmpiaurora}, including MPI collectives (e.g., \texttt{MPI\_Alltoall}) and GPCNet~\cite{GPCNet}, to identify low-performing nodes or unstable links. In the large-scale benchmarking runs, hybrid collective/P2P communication strategies (Section~\ref{sec:SoftwareOptimizations}) reduce sensitivity to localized disruptions by limiting the scope of their impact.

Despite these measures, network variability remains an inherent challenge at exascale. While application- and system-level strategies reduce the impact of these events, occasional stalls and throughput fluctuations are unavoidable when millions of components must communicate synchronously for hours at a time.

\section{Conclusion}\label{Conclusion}

This paper presented a large-scale performance-engineering and characterization study of HPL and HPL-MxP on Aurora. With validated configurations developed across successive benchmarking campaigns, Aurora sustained 1.012~EF/s in FP64 HPL and 11.64~EF/s in mixed-precision HPL-MxP on more than 9,000 nodes.

Beyond the benchmark results, the study highlights the importance of communication configuration, deterministic and locality-aware resource mapping, explicit CPU--GPU overlap, and end-to-end mixed-precision orchestration, all central to sustaining performance across multi-hour runs at near full-system scale.

Some of these lessons are Aurora-specific, while others, including locality-aware execution, communication-path matching, pipeline overlap, collective sensitivity, and hardware-informed mixed-precision design, appear broadly relevant to tightly coupled heterogeneous systems at extreme scale.
\section*{Acknowledgment}

The authors would like to thank colleagues Prem Sakarda, Nitin Dhamankar and Vitali Morozov for their valuable contributions to this work.

This research used resources of the Argonne Leadership Computing Facility, a U.S. Department of Energy (DOE) Office of Science user facility at Argonne National Laboratory and is based on research supported by the U.S. DOE Office of Science-Advanced Scientific Computing Research Program, under Contract No. DE-AC02-06CH11357.

AI-generated text assistance was used during the preparation of this manuscript. Specifically, an AI writing assistant (Anthropic Claude) was used to help refine and improve the clarity and presentation of the text in the paper. All technical content, experimental results, and scientific claims were produced entirely by the authors. The authors reviewed and verified all AI-assisted text for accuracy.

Disclaimer: Performance results are based on testing as of dates shown in configurations and may not reflect all publicly available updates. See configuration disclosure for details. No product or component can be absolutely secure.

\bibliographystyle{IEEEtran}
\bibliography{references}

\end{document}